\documentclass[useAMS,usenatbib,usegraphicx]{mn2e}
\title{On a systematic bias in SBF-based distances due to gravitational microlensing}

\author[Tuntsov \& Lewis]
{A.V. Tuntsov\thanks{E-mail: tyomich@physics.usyd.edu.au} 
\& G.F. Lewis \thanks{E-mail: gfl@physics.usyd.edu.au} \\
A29, School of Physics, University of Sydney, NSW 2006, Australia}
\begin{document}
\date{Accepted 2006 June 27. Received 2006 June 26; in original form 2006 February 20.}
\pagerange{\pageref{firstpage}--\pageref{lastpage}} \pubyear{2006}
\maketitle
\label{firstpage}
\begin{abstract}
The effect of gravitational microlensing on the determination of extragalactic distances using surface brightness fluctuations (SBF) technique is considered 
and a method to calculate SBF amplitudes in presence of microlensing is presented. With a simple approximation for the magnification power spectrum at
low optical depth the correction to the SBF-based luminosity distance is calculated. The results suggest the effect can be safely neglected at present
but may become important for SBF-based Hubble diagrams at luminosity distances of about $1\,\mathrm{Gpc}$ and beyond.
\end{abstract}
\begin{keywords}
gravitational lensing -- methods: analytical -- stars: statistics -- galaxies: distances and redshifts -- galaxies: stellar content -- dark matter
\end{keywords}
\section{Introduction}

For many centuries, measuring distances to celestial objects has been a central part of astronomy. Historically, building on the trigonometric parallax
technique, astronomers have developed a long cosmic distance ladder, within which every step is supplied by observing more and more luminous objects further
and further away, whose calibration is based on distances obtained at preceding steps.

About two decades ago a new method to determine extragalactic distances using surface brightness fluctuations (SBFs) was proposed by \citet{tonryschneider}. 
The method relies on a simple but powerful idea that less numerous stellar populations show greater relative fluctuations in their total brightness 
as Poissonian fluctuations in the number of stars are more pronounced in them compared to more numerous ones. This allows one to estimate the true average
number of stars in and therefore intrinsic luminosities of pixels in images of distant galaxies. Comparing these estimates with the observed flux then
estimates the luminosity distance directly. Although the idea of the method was clear to some astronomers long ago (see comments in \citealt{tonryschneider}), 
it was not until the mass introduction of linear panoramic detectors (CCDs) into astronomical observations, that the method could be usefully employed.

Since then, SBF technique has proven extremely fruitful in independent distance determination for galaxies and also in studies of their stellar populations
(for review material, see \citealt{blakeslee98} or \citealt{blakeslee01}). Various ways for the calibration of surface brightness fluctuations -- both 
empirically and on the basis of stellar evolution simulations -- have been developed, allowing to reduce uncertainty in distance determination to the level
of five to ten per cent, making it one of the most precise methods in extragalactic astronomy. The range of the method has also increased nearly ten-fold 
compared to original estimates, and the distances to galaxies at nearly 200 Mpc have been successfully measured using the {\it Hubble Space Telescope} 
\citep{jensen01}. 

Since the major limiting factor in the implementation of the method is photon noise, the trend in SBF observations has been towards the
infrared for at least a decade, as one gets more infrared photons at a given flux level and the SBF signal is dominated by late-type giant stars. In addition,
IR surface brightness fluctuations turned to be very interesting for stellar populations studies (e.g., \citealt{jensen03, mouhcinegonzalezliu}).

One of the external factors which can potentially contribute to the fluctuations of stellar magnitudes is gravitational microlensing by compact objects in the Universe. The presence of such population has been inferred from long-term photometrical monitoring of quasars \citep{hawkinsnature, hawkins96}. The effect of such population on photometrical and spectroscopical properties of quasars has been considered previously \citep{canizares82, dalcanton94} and useful observational constraints on the properties of such population have been obtained \citep{schneider93}.

This effect has been proposed as a tool to detect dark objects in galaxy clusters by observing {\it temporal} variations in the flux
of a given pixel \citep{liw, lewisibata, ourcluster}. However, microlensing also introduces extra scatter in flux variations {\it among different pixels} 
and makes the distances inferred via SBF method systematically lower than the true ones. The correction due to microlensing increases with redshift 
dramatically, and the compact objects therefore have a potential to profoundly affect SBF-based Hubble diagrams at high $z$.

In the present paper, we investigate how strong this effect is and whether it represents a serious issue for the present-day SBF observations. Borrowing some 
technique from our previous work on galaxy clusters \citep{ourcluster}, we show how microlensing can be taken into account for SBF studies, and then calculate relevant
quantities under some general assumptions about potential microlensing population.

We start in Section 2 with an alternative derivation of the SBF amplitude which allows to take microlensing into account. Section 3 presents a simple summation
approximation we use for the magnification spatial power spectrum and discusses a possible contribution to microlensing by 
an occasional intervening low surface brightness galaxy on the line of sight to the source. Then, in section 4 we first show that the interstellar 
correlations within the pixel can be neglected and then introduce and calculate the correction to SBF distances due to microlensing for simulated stellar 
populations. Section 5 discusses our results.

\section{Surface brightness fluctuations in presence of microlensing}

The method to estimate extragalactic luminosity distances $d$ using surface brightness fluctuations (SBF) aims at measuring the ratio between the dispersion and average 
value of the flux at the pixels comprising a distant galaxy image
\begin{equation}
\frac{\sigma^2_F}{\langle F\rangle}=\frac{1}{4\pi d^2}\frac{\sigma^2_L}{\langle L\rangle} \label{sigmaftof} ,
\end{equation}
where $L$ is the total luminosity of the pixel. For a given stellar population of the galaxy, the second fraction on the right-hand side of this equation
can be calculated using stellar evolution models. Alternatively, one can assume that this quantity does not vary much from one galaxy to another once 
they have similar stellar populations in terms of age and metalicity and determine its value by calibrating it on a sample of nearby galaxies for which
the distance $D$ can be obtained via other methods.

Any particular implementation of the method needs to find a way to estimate the intrinsic value of $\sigma^2_F/\langle F\rangle$ from an observed CCD image 
of the galaxy getting rid of different sources of noise present in the real image in comparison to an ideal one -- non-independence of the pixels introduced 
by a finite-width point-spread function of the optical system, contamination due to globular clusters of the target galaxy, photon count fluctuations and 
readout noise, sky background etc.. Advanced methods to do so have been developed over two decades which have passed since the method was proposed by 
Tonry and Schneider (1988) and the intrinsic -- in the sense that they are unaffected by the sources of noise just mentioned -- fluctuations in the flux 
can now be measured with high precision.

However, the interpretation of the quantity $L$ has always been that it is the true luminosity of the stars comprising a pixel, i.e. the sum of 
the luminosities of all stars the pixel contains. Such an interpretation does not take into account that the observed flux contributed by a star
can in fact be affected by microlensing en route to the observer. In this paper we investigate whether this is an observationally important issue 
and in which ways microlensing affects distance determination. Since $\sigma^2_F/\langle F\rangle$ can be determined from observations and there is such a simple relation~(\ref{sigmaftof}) 
between this ratio, distance $d$ and $\sigma^2_L/\langle L\rangle$, we will focus on the latter quantity and the way these apparent pixel luminosity fluctuations 
are affected by microlensing.

In the method establishing paper of Tonry and Schneider (1988), $\sigma^2_L/\langle L\rangle$ was computed in a `filling number' fluctuations manner. That is, it
was assumed that there is a certain number of different luminosity levels $l_k$ (minuscule letter $l$ is used for individual star luminosities while the
capital $L$ is reserved for the total luminosity of the pixel), and the number $N_k$ of stars of a certain type $k$ in the pixel is a Poissonian 
random variable with the mean parameter $\bar{N}_k=\bar{N}\phi_k$, where $\bar{N}$ is the mean number of stars in the pixel and $\phi_k$ is the luminosity distribution.
This method is not applicable when microlensing is involved and we will be forced to take a less elegant approach, assuming instead that luminosity of 
a given star is a random variable with distribution $\phi_k$, and the total number of stars in the pixel is a Poissonian variable; microlensing magnification
value will also be introduced as a random function. It is therefore reasonable to compare the two types of calculations and point out assumptions of the
original methods which do not hold when microlensing is involved.

The original gives the following value for the total luminosity of the pixel:
\begin{equation}
L=\sum\limits_k\, l_k N_k \label{pmethod}. 
\end{equation}
To find its average value one averages the filling numbers:
\begin{eqnarray}
\lefteqn{\langle L\rangle = \left\langle\sum\limits_k\,l_k N_k\right\rangle=\sum\limits_k\,l_k\langle N_k\rangle=\sum\limits_k\,l_k \bar{N}_k=\bar{N}\langle l\rangle_\phi } \label{averagepmethod} 
\end{eqnarray}
where $\langle l\rangle_\phi$ is the average individual star luminosity determined by its distribution $\phi$. The square of this quantity is
\begin{eqnarray}
\lefteqn{\langle L\rangle^2= \sum\limits_k\, \left(l_k\right)^2 \langle N_k\rangle^2 +\sum\limits_{k\not=k'} l_kl_{k'} \langle N_k\rangle\langle N_{k'}\rangle   } \label{squaredaveragemethodp} 
\end{eqnarray}
The average of the luminosity squared $L^2$ is, similarly,
\begin{eqnarray}
\lefteqn{\langle L^2\rangle= \left\langle\sum\limits_k \left(l_k N_k\right)^2 + \sum\limits_{k\not=k'} l_kl_{k'}N_kN_{k'}\right\rangle} \label{averagesquarepmethod} \\
& & = \sum\limits_k \left(l_k\right)^2 \left\langle N_k^2\right\rangle + \sum\limits_{k\not=k'} l_k l_{k'} \left\langle N_k N_{k'}\right\rangle \nonumber
\end{eqnarray}
and for the dispersion $\sigma^2_L=\langle L^2\rangle - \langle L\rangle^2$ one has
\begin{eqnarray}
\lefteqn{\sigma^2_L=\sum\limits_k \left(l_k\right)^2 \left(\left\langle N_k^2\right\rangle - \left\langle N_k\right\rangle^2\right) }\label{dispersionmethodp} \\
& &  + \sum\limits_{k\not=k'} l_kl_{k'}\left(\left\langle N_k N_{k'}\right\rangle - \left\langle N_k \right\rangle\left\langle N_{k'}\right\rangle\right) \nonumber
\end{eqnarray}
The value in the brackets on the first line of the right-hand side is, for Poissonian distribution, $\sigma^2_{N_k}=\bar{N}_k=\bar{N}\phi_k$ while
the second line vanishes as $\langle N_k N_{k'}\rangle =\langle N_{k'}\rangle \langle N_{k'}\rangle $ because the numbers of different stars in the pixel 
are independent -- at least, this is a very reasonable approximation if the galaxy stellar population is considered homogeneous. As a result, one has
\begin{equation}
\sigma^2_L=\bar{N}\sum\limits_k \left(l_k\right)^2\phi_k=\bar{N}\left\langle l^2\right\rangle_\phi \, . \label{methodpdispersionresult}
\end{equation}
Since Poissonian statistics remains valid for arbitrary low mean numbers $\bar{N}_k$, the transition to continuous distribution density $\phi(l)$ requires
no effort.

The independence mentioned above is lost when microlensing is introduced -- the magnification values $\mu$ inside the pixel are correlated: knowledge that
a certain number $N_j$ of stars have magnification value $\mu_j$ can be used to predict the probability that the number of stars with magnification $\mu_{j'}$ 
is $N_{j'}$. We therefore have to resort to a different solution.

In an alternative approach, the values considered random are
\begin{itemize}
\item the total number of stars in the pixel $N$, its distribution is assumed Poissonian with mean $\bar N$;
\item the type of each star, characterized by its luminosity $l_i$ and spatial profile $\psi_i({\bf r})$ ($i=\overline{1, N}$); 
the values of the luminosity $l$ will be randomly drawn with the distribution $\phi(l)$ independently of each other\footnote{For simplicity, we assume a unique
relation between $l_i$ and $\psi_i$ although index $i$ can denote both without any changes to the notation.};
\item the coordinates of each star within the pixel ${\bf r}_i$ with $i=\overline{1, N}$, these coordinates will be considered independently and uniformly distributed in the pixel;
\item the magnification field realization $\mu({\bf r})$, its properties will be specified as we progress through the calculation.
\end{itemize}
When calculating $\sigma^2_L$ we will have to average with respect to all of these quantities, and, where necessary, appropriate averages will be denoted by 
$N$, $\phi$, ${\bf r}$ or $\mu$ subscripts at the angled brackets, respectively.

Then, the apparent luminosity of a certain realization of the pixel is
\begin{equation}
L=\sum\limits_i^N l_i\mu_i({\bf r}_i) \label{omethod}.
\end{equation}
As the magnification can depend on the type of the source star -- most importantly, its size, --  $\mu$ was given the index $i$ as well. 
These
values can be calculated by convolving $\psi_i({\bf r})$ and the underlying magnification map $\mu({\bf r})$ and therefore $\mu_i$ Fourier transform is the
product of $\psi_i$ and $\mu$ transforms (times $2\pi$ as we choose Fourier transformation kernel normalization 
in symmetric convention such that only the phase is reversed in the inverse transform):
\begin{equation}
\tilde\mu_i({\bf k})=\tilde\mu({\bf k})2\pi\tilde\psi_i({\bf k}) \label{muFourier}
\end{equation}
In particular, for a statistically homogeneous magnification map the average does not depend on $\bf r$:
\begin{equation}
\langle\tilde\mu({\bf k})\rangle_\mu=\langle\mu\rangle 2\pi\delta({\bf k}) , \label{muFourierdelta}
\end{equation}
where $\langle\mu\rangle$ is the average magnification. Therefore,
\begin{equation}
\langle\mu_i\rangle_\mu=\langle\mu\rangle_\mu 2\pi\tilde\psi_i(0)=\langle\mu\rangle_\mu .\label{muiaverage}
\end{equation}
since $2\pi\tilde\psi_i(0)=1$ -- it is more convenient to put the normalization of the spatial profile into the value of $l_i$. Then, one has
\begin{equation}
\langle L\rangle = \left\langle\sum\limits_{i=1}^N \left\langle l_i\langle\mu\rangle_\mu\right\rangle_\phi\right\rangle_N = \bar{N}\langle\mu\rangle_\mu\langle l\rangle_\phi  . \label{omethodaverageLmuphi}
\end{equation}
The square of the pixel luminosity is
\begin{equation}
L^2=\sum\limits_i^N l_i^2\mu_i^2({\bf r}_i) + \sum\limits_{i\not=i'}^N l_i l_{i'}\mu_i({\bf r}_i)\mu_{i'}({\bf r}_{i'}) .\label{omethodLsquared}
\end{equation}
One first averages this over different $\mu({\bf r})$ realizations
\begin{equation}
\left\langle L^2\right\rangle_\mu=\sum\limits_i^N l_i^2\langle\mu_i^2({\bf r}_i)\rangle_\mu +  \sum\limits_{i\not=i'}^N l_i l_{i'}\langle\mu_i({\bf r}_i)\mu_{i'}({\bf r}_{i'})\rangle_\mu .  \label{omethodLsquaredmu} 
\end{equation}
Homogeneity of the underlying magnification map $\mu({\bf r})$ implies that its autocorrelation function depends only on the difference of its arguments:
\begin{equation}
\langle\tilde\mu({\bf k}_1)\tilde\mu({\bf k}_2)\rangle_\mu=P_\mu({\bf k}_1)\delta({\bf k}_1+{\bf k}_2) , \label{secondorderhomogeneity}
\end{equation}
where $P_\mu({\bf k})$ is the magnification spatial power spectrum, or $2\pi$ the Fourier transform of the autocorrelation function. 
It follows then that the factor in front of a $\delta$-function in the Fourier transform of the cross-correlation between the magnifications 
of stars $i$ and $i'$ is
\begin{equation}
P_{ii'}({\bf k})=P_\mu({\bf k})\,4\pi^2\tilde\psi_i({\bf k})\tilde\psi_{i'}(-{\bf k}) \label{mixedcorrelations}
\end{equation}
and, in particular, the diagonal values of the cross-correlation matrix are independent of ${\bf r}$:
\begin{equation}
\langle\mu^2_i({\bf r})\rangle_\mu=\int\mathrm{d}^2{\bf k}\,P_\mu({\bf k})|\tilde\psi_i({\bf k})|^2=\langle\mu^2_i\rangle_\mu .\label{averagemusquared}
\end{equation}
Averaging with respect to ${\bf r}_i$ is performed by convolving~(\ref{omethodLsquaredmu}) with the distribution of ${\bf r}_i$, $i=\overline{1, N}$. 
Since these coordinates are assumed independent and uniformly distributed in the pixel $\mathcal{S}$ of area $||\mathcal{S}||$, this leads to
\begin{equation}
\left\langle L^2\right\rangle_{\mu, {\bf r}_i} = \sum\limits_i^N l_i^2\langle\mu_i^2\rangle_\mu + \sum\limits_{i\not=i'}^N l_il_{i'}\langle\mu_i\mu_{i'}\rangle_{\mathcal{S}}, \label{omethodLsquaredmur}
\end{equation}
where the average value of the correlation function
\begin{equation}
\langle\mu_i\mu_{i'}\rangle_{\mathcal{S}}\equiv||\mathcal{S}||^{-2}\int\limits_{\mathcal{S}^2}\mathrm{d}^2{\bf r}_1\mathrm{d}^2{\bf r}_2\,\langle\mu_i({\bf r}_1)\mu_{i'}({\bf r}_2)\rangle_\mu \label{averagecorrelationdef}
\end{equation}
can be computed using~(\ref{mixedcorrelations}):
\begin{equation}
\langle\mu_i\mu_{i'}\rangle_{\mathcal{S}}=\int\mathrm{d}^2{\bf k}\,P_\mu({\bf k})\tilde\psi_i({\bf k})\tilde\psi_{i'}^*({\bf k}) |2\pi\tilde{s}({\bf k})|^2 .\label{mixedmuaverage}
\end{equation}
Here $\tilde{s}({\bf k})$ is the coordinates characteristic function, or the Fourier transform of its distribution density $s({\bf r})$:
\begin{equation}
s({\bf r})=||\mathcal{S}||^{-1}\left[\matrix{1, \hspace{1cm} {\bf r}\in\mathcal{S} \cr 0, \hspace{1cm} {\bf r}\not\in\mathcal{S}}\right. \label{pixel}
\end{equation}

Since each star type is chosen independently of all others, one has
\begin{equation}
\langle l_i^2\langle\mu_i^2\rangle_\mu\rangle_\phi=\langle l^2\rangle_\phi\int\mathrm{d}^2{\bf k}\,P_\mu({\bf k})|\langle\tilde\psi^2({\bf k})\rangle_\phi| \label{squaredmutypeaverage}
\end{equation}
and
\begin{equation}
\left\langle l_i l_{i'}\langle\mu_i\mu_{i'}\rangle_{\mathcal{S}}\right\rangle_\phi=\langle l\rangle^2_\phi\int\mathrm{d}^2{\bf k}\, P_\mu({\bf k})|\langle\tilde\psi({\bf k})\rangle^2_\phi| |2\pi\tilde{s}({\bf k})|^2 , \label{mixedmutypeaverage}
\end{equation}
where the `luminosity-weighed' averages of the source star profile $\langle\tilde\psi({\bf k})\rangle_\phi$ and its square $\langle\tilde\psi^2({\bf k})\rangle_\phi$ have been introduced as, respectively,
\begin{equation}
\langle\tilde\psi({\bf k})\rangle_\phi\equiv \langle l\rangle_\phi^{-1}\langle l_i\tilde\psi_i({\bf k})\rangle_\phi \label{averagepsi}
\end{equation}
and
\begin{equation}
\langle\tilde\psi^2({\bf k})\rangle_\phi\equiv \langle l^2\rangle_\phi^{-1}\langle l^2_i\tilde\psi^2_i({\bf k})\rangle_\phi. \label{averagepsisquared}
\end{equation}
The number of equal terms in the first and second sums of~(\ref{omethodLsquaredmur}) is $N$ and $N(N-1)$, and their Poissonian averages
are $\bar{N}$ and $\bar{N}^2$, respectively. As a result, one has for the dispersion of $L$:
\begin{equation}
\sigma_L^2=\langle L^2\rangle - \langle L\rangle^2=\bar{N}\langle l^2\rangle_\phi\langle\mu^2\rangle_{\phi} + \bar{N}^2\langle l\rangle_\phi^2\left(\bar{P}_\mu^{\psi\mathcal{S}} - \langle\mu\rangle_\mu^2\right)  \label{omethoddispersion}
\end{equation}
with $\langle\mu^2\rangle_\phi$ and $\bar{P}_\mu^{\psi\mathcal{S}}$ defined by the integrals on the right-hand side of~(\ref{squaredmutypeaverage}) and~(\ref{mixedmutypeaverage}), respectively.

We can change the second term of the last equation into a more intuitively clear form by noticing that $\tilde\psi(0)=(2\pi)^{-1}$ and $2\pi\tilde{s}(0)=1$ 
while Fourier transform of a constant is proportional to a $\delta$-function. As a result, the value in the brackets can be calculated as
\begin{equation}
\bar{P}_{\Delta\mu}^{\psi\mathcal{S}}\equiv\bar{P}_\mu^{\psi\mathcal{S}}-\langle\mu\rangle_\mu^2=\int\mathrm{d}^2{\bf k}\,P_{\Delta\mu}({\bf k})|\langle\tilde\psi({\bf k})\rangle^2||2\pi\tilde{s}({\bf k})|^2, \label{PdeltamupsiSdef}
\end{equation}
where $P_{\Delta\mu}({\bf k})=P_\mu({\bf k})-4\pi^2\langle\mu\rangle_\mu^2\delta({\bf k})$ is the power spectrum of the magnification fluctuations $\Delta\mu({\bf r})=\mu({\bf r})-\langle\mu\rangle_\mu$.

The dispersion and average value of the observed flux in the pixel $F$ differ from those of the luminosity by factors $(4\pi d^2)^{-2}$ and $(4\pi d^2)^{-1}$, respectively. 
In the absence of microlensing -- $\langle\mu\rangle=1$, $\langle\mu^2\rangle=1$, $\bar{P}^{\psi\mathcal{S}}_{\Delta\mu}=0$ -- this gives the following 
estimate for the luminosity distance $\hat{d}$:
\begin{equation}
\hat{d}^{-2}=4\pi\frac{\sigma^2_F}{\langle F\rangle}\frac{\langle l\rangle}{\langle l^2\rangle} . \label{Destimate}
\end{equation}
By comparing this formula with~(\ref{omethodaverageLmuphi}) and~(\ref{omethoddispersion}) one finds that when lensing is important this estimate is biased compared to the true
distance $d$. The relationship between the two values reads:
\begin{equation}
\hat{d}^{-2}=d^{-2}\frac{\langle\mu^2\rangle_\phi}{\langle\mu\rangle}+4\pi\frac{\langle F\rangle\langle l\rangle}{\langle l^2\rangle}\frac{\bar{P}^{\psi\mathcal{S}}_{\Delta\mu}}{\langle\mu\rangle^2} \label{DhatDrelationship}
\end{equation}
or
\begin{equation}
\hat{d}^{-2}=d^{-2}\left(\frac{\langle\mu^2\rangle_\phi}{\langle\mu\rangle} + \bar{N}\frac{\langle l\rangle^2}{\langle l^2\rangle}\frac{\bar{P}^{\psi\mathcal{S}}_{\Delta\mu}}{\langle\mu\rangle} \right) .\label{DhatDotherrelationship}
\end{equation}
Since $\mu\ge 1$ and $\bar{P}^{\mu\mathcal{S}}_{\Delta\mu}\ge 0$, the estimate $\hat{d}$ is systematically lower than the true value $d$ -- microlensing introduces extra scatter in the pixel fluxes; in addition,
mass along the line of sight makes distant sources generally brighter on average (the first term on the right-hand side of $\langle\mu^2\rangle/\langle\mu\rangle=\langle\mu\rangle+\langle\Delta\mu^2\rangle/\langle\mu\rangle	$), but this magnification effect is less important than the additional scatter (see discussion in section 4).

The second term in~(\ref{DhatDrelationship}-\ref{DhatDotherrelationship}) is the most troublesome as it could potentially change the recipe for $\sigma^2_L/\langle L\rangle$
empirical determination -- the ratio between the local fluctuation amplitude and mean would no longer be constant and the way the mean is subtracted would
have to be changed. However, as we demonstrate below, typical correlation scale of the magnification map is much smaller than the typical distance between the
stars. Hence, the correlation term is small.

Further to this point, the calculation presented above does not take into account any correlations which can arise between different pixels; taking these correlations into account
would force us to significantly extend this simple study by reconsidering the way in which the intrinsic fluctuations in the flux are related to the observed CCD measurement statistics. 
However, interpixel correlations are much weaker than interstellar ones and can thus be safely neglected.

Finally, another important thing that should be addressed is the convergence of the variance. Here we only calculate the dispersion of the magnification 
value distribution, which is an overestimate of the most likely variance of a finite sample of $N_s$ stars forming the galaxy. This happens because, for 
a small source, the dispersion is dominated by a long $\propto\mu^{-3}$ tail of the distribution density \citep{peacock, schasymptotics}, which is in fact
unlikely to be properly populated by the stars. We can estimate the impact of this shortcoming by considering a model magnification distribution density
\begin{equation}
p(\mu)=\frac{2(\mu_0-1)^2(\mu - 1)}{[(\mu-1)^2+(\mu_0-1)^2]^2} \label{pmu}
\end{equation}
that takes into account three major theoretically established properties of the magnification distribution function (correct normalization and the first 
moment $\bar\mu$ and the large $\mu$ behaviour; see \citealt{ourcluster} for further details) for $\mu_0-1=2(\bar\mu-1)/\pi$. In order to model the finite 
sample effect, we, when calculating the dispersion, will truncate the integration not at the value $\mu_\mathrm{max}=2\hat{r}_E/R$, determined by 
the source size $R$ relative to the projection of the Einstein radius of the lens~(\ref{Einsteinradiusdef}), but at the value where the cumulative 
probability $\mathcal{P}(>\mu_{N_s})$ drops below the inverse size of the sample $N_s^{-1}$. This truncation occurs inside the $p(\mu)\propto\mu^{-3}$ tail 
and does not noticeably affect either the normalization or the first moment of the distribution.

In the case when the $\bar\mu-1\ll 1$, which corresponds to the low optical depth (estimates in the next two sections show that this is 
the applicable limit), the cumulative probability of the magnification greater than $\mu$ is
\begin{equation}
\mathcal{P}(>\mu)=\left[1+\left(\frac\pi2\frac{\mu-1}{\bar\mu-1}\right)^2\right]^{-1} \label{cumulativep}
\end{equation}
and therefore for $\mu_{N_s}$ where $\mathcal{P}(>\mu_{N_s})$ drops below $N_s^{-1}$ one has
\begin{equation}
\mu_{N_s}=1+\frac{2}\pi(\bar\mu-1)\sqrt{N_s-1} . \label{muNs}
\end{equation}
At redshift $z\sim 0.1$, the average magnification value due to a population of lenses that comprise a fraction $\Omega_d$ of the critical density,
is roughly $\bar\mu-1\sim 10^{-3}\Omega_d$ (see figures~\ref{qfig}, \ref{qlfig}). Therefore, for a target galaxy with $N_s\sim 10^{11}$ stars in it, the bound 
of $\mu$ due to the sample size is $\mu_{N_s}\sim (10^2 - 10^3)\Omega_d$. At the same time, the other bound, which is due to the finite size of the source
is, for a Solar-mass lens, $\mu_\mathrm{max}\sim 10^{3} - 10^{6}$, depending on the size of the source star. Given that the dispersion is roughly 
proportional to the logarithm of the truncation value $\mu$ (since $p(\mu)\mu^2\propto\mu^{-1}$ at large $\mu$), the finite sample effect can decrease 
the value of the microlensing correction to SBF-based distances by a factor of a few. 

This reduction of the effect magnitude is dependent on the 
number of stars in the galaxy and will not be calculated beyond this estimate. As we will see in the following, the effect can be safely neglected
in present but the finite-sample reduction should be kept in mind when the point is reached observationally for the effect to be important; in such 
cases the finite-sample reduction should be calculated individually for every galaxy.

\section{Magnification power spectrum}

\subsection{The summation approximation}

In this section we calculate the magnification power spectrum $P_\mu({\bf k})$ using an approximation that magnifications due to individual lenses
simply add up in the limit of very low optical depth.

This is different from the widely used multiplication approximation \citep{vietriostriker, pei, schneider93} where the magnification $\mu$ 
of a source seen through a number of lensing planes is taken as the product of the magnification values $\mu_i$ each lensing plane has 
at the position of the source ${\bf r}_s$:
\begin{equation}
\mu({\bf r}_s)\approx\prod\limits_i \mu_i({\bf r}_s) , \label{multiplicationapproximation}
\end{equation}
where the product is over all lensing plane along the line of sight.

We, instead, assume that this value can be calculated as a sum:
\begin{equation}
\mu({\bf r}_s) - 1 \approx\sum\limits_i \left(\mu_i({\bf r}_s)-1\right),  \label{summationapproximation}
\end{equation}
where the summation extends over all lenses. It is much simpler in this approximation to calculate the average and the power spectrum of the magnification
due to the linearity of taking the average.

This assumption could be justified by saying that the two approaches give very  similar results when all but one of the magnifications are close to unity 
and a reference to the authority of the multiplication approximation. We, however, feel the need to go into a little more detail to explain the physics
behind this assumption. 

In order to do so, one can consider the morphology of a macroimage of which a classic example is given in a figure by Paczy\'nski (1986).
Generally, when there are $N$ (point-like) microlenses in the field, one has at least $N+1$ images of which one -- the primary -- is located close
to the unperturbed direction to the source and the other $N$ secondaries are formed near the lenses. These latter microimages are highly demagnified unless the line of sight
to the source passes close to the Einstein-Chwolson ring of one of the lenses in which case the magnification gets larger than $3/\sqrt{5}\approx 1.34$.
The fluxes from all microimages add up.

When lenses are randomly distributed in space, the average number of lenses within their Einstein radii of the line of sight is called
the optical depth:
\begin{equation}
\tau\equiv\int\mathrm{d}D\,\pi r_E^2(D) n(D), \label{taudef}
\end{equation}
where integration extends from the observer to the source, $n(D)$ is the number density of lenses and the Einstein radius of a lens of mass
$M$ at the angular diameter distance $D_L$ from the observers and $D_{LS}$ from the source is
\begin{equation}
r_E=\sqrt{\frac{4GM}{c^2}\frac{D_L D_{LS}}{D_S}} .\label{Einsteinradiusdef}
\end{equation}

For a cosmological population of compact lenses with constant closure density fraction $\Omega_d$ in a flat Universe, one has, at low redshifts $z\la 0.1$ (e.g. \citealt{pressandgunn}):
\begin{equation}
\tau\approx\int\limits_0^{D_s}\mathrm{d}D\,\frac{4\pi GMn}{c^2}\frac{D(D_s-D)}{D_s}=\Omega_d\frac{z^2}{4} \label{tauestimate}
\end{equation} 
and thus is relatively small.

This quantity is also a measure of how many microimages will be appreciably magnified since the Poissonian statistics implies 
that this number is $\tau\mathrm{e}^{-\tau}$ which is approximately $\tau$ when $\tau\ll 1$; we consider just this case of low optical depth in the present study. 
A situation where more than one lens contributes noticeable flux to the macroimage is less likely by a factor of approximately $3\tau/2$.

In addition to always present $N+1$ image, additional pairs of (tertiary?) images may form when the source moves into a region bound by a caustic curve.
Magnification diverges at these curves so these additional images could contribute significantly to the total flux. However, Wambsganss, Witt and Schneider
(1992) derived the mean number of these extra pairs to be $\bar{N}_p\approx\tau^2$ in the case of zero external shear which is also very low. Granot, Schechter and
Wambsganss (2003) calculate this quantity for non-zero shear numerically and find similar behaviour at low values of $\tau$; they also show that the 
statistics of $N_p$ is approximately Poissonian, so that $\tau\ll 1$ ensures these images can be neglected on average. These studies relate to 
two-dimensional distribution of lenses and there have not been studies exploring a 3D case but at low optical depth there should not be much difference.

Each of the secondary microimages contributes $\propto(\mu_i-1)/2$ to the total flux and shears the beam of the primary image in such a way that 
its area, on average, shrinks by a fraction $(\mu_i-1)/2$. As a result, the input of every lens to the magnification value is approximately $\mu_i-1$ and 
the total magnification is given by~(\ref{summationapproximation}). 

Certainly, (\ref{multiplicationapproximation}) and~(\ref{summationapproximation}) agree in the limit of average individual magnification values close 
to unity. It is also trivial that~(\ref{multiplicationapproximation}) can be written as a sum simply by applying logarithm to both sides but averaging this
quantity over possible magnification map configurations would produce statistical properties of $\ln\mu$ which are hard to convert to those of $\mu$ 
unless some additional assumptions are invoked.

\subsection{Summation for randomly distributed lenses}

We calculate the average magnification and the magnification power spectrum in the approximation~(\ref{summationapproximation}) assuming that lenses
are distributed uniformly and independently of each other.

Consider a cylinder of cross-section $\mathcal{L}$ between the observer at $D=0$ and the source at $D=D_s$ with $N$ lenses inside it. For a point-like lens of mass $M$ at
position ${\bf r}_i\in\mathcal{L}$, $D_i$, individual magnification value is given by (e.g., Schneider, Ehlers and Falco 1992):
\begin{equation}
\mu_i({\bf r}_s)-1 =\frac{y_i^2+2}{y_i\sqrt{y_i^2+4}}-1, \label{individualmagnification}
\end{equation}
where
\begin{equation}
{\bf y}_i=\frac{1}{\hat{r}_{Ei}}\left({\bf r}_s-\frac{D_s}{D_i}{\bf r}_i\right) \label{yidef}
\end{equation}
and the Einstein radius projection onto the source plane is
\begin{equation}
\hat{r}_{Ei}=\frac{D_s}{D_i}r_{Ei}=\sqrt{\frac{4GM_i}{c^2}\frac{D_s(D_s-D_i)}{D_i}}, \label{redef}
\end{equation}
where we have assumed that the distance between the $i$-th lens and the source is $D_{LS}=D_s-D_i$, which is a good approximation for
source at low redshift $z\la 0.1$ in a spatially flat Universe.

Then for the Fourier transform of $\nu\equiv\mu-1$ is given by the sum
\begin{equation}
\tilde\nu({\bf k})=\sum\limits_i^N\tilde\nu_i({\bf k}) \label{nuFouriertransform}
\end{equation}
and each term amounts to
\begin{eqnarray}
\lefteqn{\tilde\nu_i({\bf k})=\int\limits_{\mathrm{R}^2}\mathrm{d}^2{\bf r}_s\,\frac{\exp[-\mathrm{i}{\bf k}\cdot{\bf r}_s]}{2\pi}\left(\frac{y_i^2+2}{y_i\sqrt{y_i^2+4}}-1\right) } \label{nuiFouriertransform} \\
& &  = \exp\left[-\mathrm{i}{\bf k}\cdot{\bf r}_i\frac{D_s}{D_i}\right] \hat{r}^2_{Ei}\,\tilde\nu_0(\hat{r}_{Ei}{\bf k}), \nonumber
\end{eqnarray}
where we changed to the integration variable ${\bf r}_i\to {\bf y}_i$ according to~(\ref{yidef}) and introduced a dimensionless Fourier transform
\begin{eqnarray}
\lefteqn{\tilde\nu_0(\bxi)\equiv\int\limits_{\mathrm{R}^2}\mathrm{d}^2{\bf y}\,\frac{\exp[-\mathrm{i}\bxi\cdot{\bf y}]}{2\pi}\left(\frac{y^2+2}{y\sqrt{y^2+4}}-1\right)} \label{nu0def}\\
& & =\int\limits_0^\infty\mathrm{d}y\,yJ_0(\xi y)\left(\frac{y^2+2}{y\sqrt{y^2+4}}-1\right)=\tilde\nu_0(\xi) . \nonumber
\end{eqnarray}
This function is $\approx\xi^{-1}$ for large $\xi$ and tends to unity as $\xi\to 0$; a good approximation can be provided by
\begin{equation}
\tilde\nu_a(\xi)=\frac{1}{\sqrt{\xi^2+1}} \label{nuadef}
\end{equation}
which only slightly (by less than two per cent compared to the numerically evaluated $\tilde\nu_0(\xi)$) overestimates its value at low $\xi\la 1$. 
%
%
%
%

Averaging with respect to the position of lens in the appropriate plane ${\bf r}_i\in\mathcal{L}$ produces, in the limit of infinite cylinder cross-section $\mathcal{L}\to\mathcal{R}^2$
\begin{equation}
\left\langle\exp\left[-\mathrm{i}{\bf k}\cdot{\bf r}_i\frac{D_s}{D_i}\right]\right\rangle_{{\bf r}_i}=\frac{(2\pi)^2}{||\mathcal{L}||}\left(\frac{D_i}{D_s}\right)^2\delta({\bf k})+\overline{o}\left(||\mathcal{L}||^{-1}\right) \label{averageoverri}
\end{equation}
and if the distribution of lenses along the $D$ coordinate is some $\varphi_D(D)\propto n(D)$, one has in $\tilde\nu({\bf k})$ a sum of $N$ equal terms:
\begin{equation}
\langle\tilde\nu_i({\bf k})\rangle_\mu=\frac{2}{||\mathcal{L}||}\frac{2\pi\delta({\bf k})}{\int\mathrm{d}D\,n(D)}\int\limits_0^{D_s}\mathrm{d}D\,\pi r^2_{E}(D) n(D)\tilde\nu_0(\hat{r}_{E}{\bf k}). \label{nuifinal}
\end{equation}
Since $\tilde\nu_0(0)=1$ and the average number of lenses within the cylinder is $\bar{N}=\int\mathrm{d}D\,||\mathcal{L}||n(D)$ we arrive at
\begin{equation}
\langle\tilde\nu({\bf k})\rangle_\mu=\langle\nu\rangle_\mu2\pi\delta({\bf k})=\left(\langle\mu\rangle_\mu-1\right)2\pi\delta({\bf k})=2\tau\cdot 2\pi\delta({\bf k}). \label{nufinal}
\end{equation}
The optical depth is proportional to the first power of mass of the lens ($r_E\propto\sqrt{M}$), therefore a distribution of masses can be taken into account
by using the average value of the mass.

In a similar fashion, when calculating the correlation function, one has:
\begin{eqnarray}
\lefteqn{\tilde\nu({\bf k}_1)\tilde\nu({\bf k}_2)=\sum\limits_{i}^N\tilde\nu_i({\bf k}_1)\tilde\nu_i({\bf k}_2)+\sum\limits_{i\not=i'}^N\tilde\nu_i({\bf k}_1)\tilde\nu_{i'}({\bf k}_2)  } \label{nuk1nuk2} \\
& &\hspace{-5mm} =\sum\limits_i^N \hat{r}_{Ei}^4 \exp\left[-\mathrm{i}\left({\bf k}_1+{\bf k}_2\right)\hat{\bf r}_i\right]\tilde\nu_0\left(\hat{r}_{Ei}{\bf k}_1\right)\tilde\nu_0\left(\hat{r}_{Ei}{\bf k}_2\right) \nonumber \\
& &\hspace{-5mm} +\sum\limits_{i\not=i'}^N \hat{r}_{Ei}^2\hat{r}_{Ei'}^2 \exp\left[-\mathrm{i}{\bf k}_1\hat{\bf r}_i-\mathrm{i}{\bf k}_2\hat{\bf r}_{i'}\right]\tilde\nu_0\left(\hat{r}_{Ei}{\bf k}_1\right)\tilde\nu_0\left(\hat{r}_{Ei'}{\bf k}_2\right), \nonumber
\end{eqnarray}
where for typographical clarity we used lens coordinate projections onto the source plane: $\hat{\bf r}_i={\bf r}_i D_s/D_i$.

Averaging of the second sum produces $N(N-1)$ equal terms of the form:
\begin{equation}
\langle\tilde\nu_i({\bf k})\rangle_\mu^2=\left(\frac{2\tau}{\bar{N}}\right)^2\,4\pi^2\delta({\bf k}_1)\delta({\bf k}_1+{\bf k}_2) \label{nusquaredaverage}
\end{equation}
while the first sum produces $N$ terms of the form
\begin{eqnarray}
\lefteqn{\langle\tilde\nu_i({\bf k}_1)\tilde\nu_i({\bf k}_2)\rangle_\mu=}	 \label{diagonalmu} \\ 
& & \hspace{-7mm}=\delta({\bf k}_1+{\bf k}_2)\left\langle\frac{2\pi}{||\mathcal{L}||}\left(\frac{D_i}{D_s}\right)^2\hat{r}_{Ei}^4(D_i)\left|\tilde\nu_0\left({\bf k}_1\hat{r}_{Ei}(D_i)\right)\right|^2\right\rangle_{D_i}. \nonumber
\end{eqnarray}

Taking into account equation~(\ref{nufinal}) and the fact that the average values of $N$ and $N(N-1)$ are $\bar{N}$ and $\bar{N}^2$, respectively, 
one has
\begin{eqnarray}
\lefteqn{\langle\tilde\nu({\bf k}_1)\tilde\nu({\bf k}_2)\rangle_\mu=\delta({\bf k}_1+{\bf k}_2)\left(P_{\Delta\nu}({\bf k}_1)+4\pi^2\delta({\bf k}_1)\langle\nu\rangle_\mu^2\right),} \label{nuk1nuk2average}
\end{eqnarray}
where the power spectrum $P_{\Delta\nu}({\bf k})=P_{\Delta\mu}({\bf k})$ is
\begin{eqnarray}
\lefteqn{P_{\Delta\mu}({\bf k})\equiv\int\limits_0^{D_s}\mathrm{d}D\, 2\pi n(D)\left(\frac{D}{D_s}\right)^2\hat{r}_{E}^4(D)\left|\tilde\nu_0\left({\bf k}\hat{r}_{E}(D)\right)\right|^2 } \label{Pmudef} \\
& & \hspace{4mm} = 2\int\limits_0^{\tau(z_s)}\mathrm{d}\tau(D)\, \hat{r}_E^2(D)|\tilde\nu_0({\bf k}\hat{r}_E(D))|^2 \nonumber
\end{eqnarray}
-- that is, twice the cumulative optical depth times the differential optical depth-weighed average value of $\hat{r}_E^2|\tilde\nu_0({\bf k}\hat{r}_E)|^2$.

At low redshifts, using~(\ref{nuadef}), one can write the following approximation for a population of lenses with constant number density:
\begin{eqnarray}
\lefteqn{P_\mu({\bf k})=(1+4\tau)4\pi^2\delta({\bf k})+\frac{4\tau \hat{r}_E^2}{9(1+k\hat{r}_E+2k^2\hat{r}_E^2)} + \overline{o}(\tau) , }\label{Pmufinal}
\end{eqnarray}
valid to within four per cent for any ${\bf k}$, where the Einstein radius projection for the lens halfway to the source is
\begin{equation}
\hat{r}_E=\sqrt{\frac{4GM}{c^2}D_s} .\label{rEdef}
\end{equation}



\subsection{The contribution of LSB galaxies}

An inevitable source of noise to the microlensing correction that needs to be quantified is the potential presence of a low surface brightness (LSB) galaxy 
on the line of sight to the target galaxy. Even if the dark matter were not composed of any compact sources, such a galaxy would cause microlensing on its 
stars and as up to a half of the galaxy population is in the LSB form \citep{bothunimpeymcgaugh}, a significant fraction 
($\sim 10 - 50$ per cent) of the potential targets at high redshifts $z\ga 1$ will be lensed by these galaxies, although a confident estimate is difficult 
to make because of our still poor knowledge of the density and clustering of galaxies at the low end of the surface brightness distribution \citep{impeybothun}.

The effect of such a galaxy on the microlensing contribution to the SBF amplitude can easily be estimated using the results of \citet{ourcluster}, where
some of the formalism of the present paper has already been presented, and can be straightforwardly borrowed to consider the effect of a two-dimensional 
lens on the fluctuation properties. Using their equation~(14) and neglecting the interstellar correlation term in~(\ref{omethoddispersion}) of the present 
paper (see below) we find that the effect of a low surface brightness galaxy on the SBF amplitude can be expressed as an extra factor of 
\begin{equation}
\frac{\langle\mu^2\rangle_\mu}{\langle\mu\rangle_\mu}=\langle\mu\rangle_\mu\varepsilon^2_\mu  ,\label{lsbeffect}
\end{equation}
where $\varepsilon^2_\mu$ is the `correlation amplitude' defined by~(19) of \citet{ourcluster}.

To determine the two factors above, we need to know the typical mass of the microlenses and calculate the dimensionless convergence $\kappa$ 
and shear $\gamma$ of their population in the foreground galaxy, as well as their `effective' values \citep{paczynskieffective, krs} given
by the mix of the compact and smoothly distributed matter in the galaxy. As for the first ingredient, the resulting value of the dispersion is very 
weakly dependent on the ratio of the microlens mass provided that the ratio of its Einstein radius to the source size is large enough 
(see the Appendix B of \citealt{ourcluster} and discussion following~(\ref{fadef}) in this paper), which is the case here, as shown in the next subsection.
We, therefore, can adopt solar values for an estimate.

Then, one has $\langle\mu\rangle_\mu=|(1-\kappa)^2-\gamma^2|^{-1}$ and $\epsilon^2_\mu(\kappa_\mathrm{eff}, \gamma_\mathrm{eff}, \hat{r}_E/R)$ calculated 
in subsection 2.2 and shown in Figure~1 of \citet{ourcluster}. The value of the scaled convergence can be calculated from the observed surface brightness
of the foreground galaxy $u$ (in magnitudes per square arcsecond) and the mass-to-light ratio $\Upsilon$ of the component of interest by comparing it with 
the bolometric irradiance of the Earth by the Sun as
\begin{eqnarray}
\lefteqn{\kappa=\frac{4\pi G M_\odot}{c^2\cdot\mathrm{AU}^2}\Upsilon\frac{D_{LS}D_{OL}}{D_{OS}}\left(\frac{1\,\mathrm{rad}}{1"}\right)^2(1+z_l)^4 10^{0.4(m_\odot^{\mathrm{bol}}-u)}} \label{kappafromsb} \\
& & \approx 2.04\times 10^{-3-0.4(u-22.5)} \Upsilon (1+z_l)^4 \frac{D_{LS}}{D_{OS}}\frac{D_{OS}}{100\,\mathrm{Mpc}} .\nonumber
\end{eqnarray}
Here $D_{LS}$, $D_{OL}$ and $D_{OS}$ are the angular diameter distances between the lens and the source, the observer and the lens and the observer and 
the source, respectively, and $z_l$ is the lens redshift. Given that the mass-to-light ratio for stellar populations are of order unity (this value is not 
known well for LSB galaxies though), and for a galaxy to be non-detectable one needs its surface brightness to be below $\sim 23^m-24^m$ per square 
arcsecond, the typical values of the scaled convergence in stars are of order $10^{-3} - 10^{-2}$ up to moderate redshifts. If, in addition, one assumes 
that a significant amount of mass in LSB galaxies is present in a dark and smooth form \citep{impeybothun}, the total convergence is 
$\kappa_{tot} \sim 10^{-2} - 10^{-1}$; the difference between the stellar and effective convergence values is insignificant in this case.

To evaluate the shear $\gamma$, one would need an accurate determination of the surface brightness density over the entire surface of the foreground galaxy,
which is not possible if it is invisible. Generally, however, the scaled shear takes the values of the same order as the scaled convergence, and 
$\epsilon^2_\mu\sim\mathcal{O}(\kappa)$ in this case (again, assuming the typical Einstein radius is much larger than the typical target star size). 
Therefore, a typical contribution to the correction of the SBF amplitude due to an occasional intervening LSB galaxy along the line of sight is 
of order $10^{-3} - 10^{-2}$ at moderate redshifts. This is less than what we find in the next section for a cosmological distribution of compact objects 
if they make up any significant fraction of the matter content of the Universe, but can be comparable and even dominant in case no such distribution is
present.

\section{Evaluation of corrections}

\subsection{Unimportance of interstellar correlations}

The expressions above are sufficient for the calculation of $\langle\mu^2\rangle_\phi$ and $\bar{P}^{\psi\mathcal{S}}_{\Delta\mu}$ once the average source 
star profiles $\langle\tilde\psi({\bf k})\rangle_\phi$, $\langle\tilde\psi^2({\bf k})\rangle_\phi$ and the pixel shape $\tilde{s}({\bf k})$ have been specified.

However, we will first make some estimates to see what the typical values of the parameters of the problem are. There are four main length scales of the
problem: the source size $R$, the Einstein radius projection to the source $\hat{r}_E$ and the pixel size $x$, as well as the typical projected separation 
of stars in the galaxy $\Delta$.

The pixel size is determined by the resolution of the CCD matrix and the distance to the source. The typical pixel size is, in the angular measure, of order
$\theta\approx 0.2\arcsec\approx 10^{-6}\,\mathrm{rad}$. At the distance $D$ this translates into:
\begin{eqnarray}
\lefteqn{x=100\,\mathrm{pc}\,D_{100}\, \theta_{0.2}=3.1\times 10^{20}\,\mathrm{cm}\,D_{100}\,\theta_{0.2}}  \label{pixelsize}\\
& & \hspace{-5mm} = 4.1\,\mathrm{kpc}\, z\,\theta_{0.2} = 1.3\times 10^{22}\,\mathrm{cm}\,z\,\theta_{0.2}, \nonumber
\end{eqnarray}
where $D_{100}=D/100\,\mathrm{Mpc}$, $\theta_{0.2}=\theta/0.2\arcsec$ and the second line is valid at low redshifts $z\la 0.1$.

The Einstein radius projection (for the lens halfway to the source) is, according to~(\ref{rEdef}),
\begin{eqnarray}
\lefteqn{\hat{r}_E=4.5\times 10^{-3}\,\mathrm{pc}\,D_{100}^{1/2}\,m^{1/2}= 1.4\times 10^{16}\,\mathrm{cm}\,D_{100}^{1/2}\,m^{1/2} }\label{rEsize}\\
& & \hspace{-3mm} =2.8\times 10^{-2}\,\mathrm{pc}\,z^{1/2}\,m^{1/2}= 8.7\times 10^{16}\mathrm{cm}\, z^{1/2}\,m^{1/2} , \nonumber
\end{eqnarray}
where $m=M/M_\odot$.

The stars which contribute non-negligibly to the luminosity all lie on the main sequence or above it in the Hertzsprung-Russell diagram. Their sizes therefore
lie in the range:
\begin{equation}
R\approx 3\times 10^{-9}\,\mathrm{pc} - 3\times 10^{-6}\,\mathrm{pc}\approx 10^{10}\,\mathrm{cm} - 10^{13}\,\mathrm{cm}. \label{Rsize}
\end{equation}
The projected density of stars in galaxies hardly exceeds $10^4\,\mathrm{pc}^{-2}$ in their densest parts and 
is normally of order $10^2\,\mathrm{pc}^{-2}$ on average. Therefore, for the mean separation of stars in the pixel one has
\begin{equation}
\Delta\ga 10^{-2}\,\mathrm{pc} - 10^{-1}\,\mathrm{pc}\approx 3\times10^{16}\,\mathrm{cm} - 3\times10^{17}\,\mathrm{cm}. \label{separationsize}
\end{equation}
Therefore, in all extragalactic cases one has $R\ll\hat{r}_E\ll x$. The respective scales in the Fourier images domain, which are given roughly
by the inverse of these quantities and represent size of the region in ${\bf k}$ plane within which $\tilde\psi$, $P_{\Delta\mu}$ and $\tilde{s}$ differ appreciably from zero
satisfy the opposite relation.

This observation allows us to interpret and estimate the second term in the brackets of~(\ref{DhatDotherrelationship}). Namely, the integral~(\ref{PdeltamupsiSdef}) is
contributed almost exclusively by the region of size $1/x$ within which the values of $P_{\Delta\mu}$ and $\tilde\psi$ are nearly constant and equal
to $4\tau\hat{r}_E^2/9$ and $1/4\pi^2$, respectively. As a result,
we can write the following estimate for the correlation term using~(\ref{Pmufinal}):
\begin{equation}
\bar{P}^{\psi\mathcal{S}}_{\Delta\mu}\approx ||\mathcal{S}||^{-1} \frac{\tau\hat{r}_E^2}{9\pi^2}. \label{Pboldguess}
\end{equation}
Therefore, $\bar{N}\bar{P}^{\psi\mathcal{S}}_{\Delta\mu}$ is proportional to the average number of stars within the Einstein-Chwolson circle of the typical lens $\bar{N}_E$.
Therefore, taking into account the estimates~(\ref{rEsize}, \ref{separationsize}), $\bar{N}_E\sim 10^{-4} - 1$ for distances from $100\,\mathrm{Mpc}$ up to a few Gpc.

Finally, for the luminosity ratio one has:
\begin{equation}
\frac{\langle l\rangle^2}{\langle l^2\rangle}=\frac{\langle l\rangle}{\bar{l}} =10^{0.4(\overline{\mathrm{M}}-\langle\mathrm{M}\rangle)} , \label{ltolbar}
\end{equation}
where, in accordance with the tradition adopted in SBF studies, $\bar{l}$ has been used to denote the luminosity-weighed average luminosity of source stars 
(this quantity is also called the surface brightness fluctuations amplitude in view of~(\ref{Destimate})); upright $\mathrm{M}$ 
are used 
for corresponding absolute magnitudes to avoid confusion with the lens mass. Due to a great range in stellar luminosities, $\overline{\mathrm{M}}$ assumes 
values close to the absolute magnitudes of red giant branch stars while $\langle\mathrm{M}\rangle$ are closer to those of red dwarfs dominating the main sequence.
The ratio of their luminosities ranges from about a hundred to thousands, depending on the band. Therefore,
\begin{equation}
\bar{N}\frac{\langle l\rangle^2}{\langle l^2\rangle}\frac{\bar{P}^{\psi\mathcal{S}}_{\Delta\mu}}{\langle\mu\rangle}\approx\frac{\tau\bar{N}_E}{9\pi^2\langle\mu\rangle}\, 10^{0.4(\overline{\mathrm{M}}-\langle\mathrm{M}\rangle)}\ll \tau \label{PpsiSdeltamuNbarestimate}
\end{equation}
The other term in the brackets of~(\ref{DhatDotherrelationship}) is given by the convolution of $P_\mu$ with the source star profile. Assuming all stars are Lambert discs of the same radius $R$, one has
\begin{equation}
\tilde\psi({\bf k})=\tilde\psi(k)=\frac{J_1(kR)}{\pi k R} \label{Lambertdisctildepsi}
\end{equation}
and~(\ref{averagemusquared}) taken with~(\ref{Pmufinal}) gives
\begin{equation}
\frac{\langle\mu^2\rangle}{\langle\mu\rangle}=\langle\mu\rangle+\frac{8\tau}{9\pi\langle\mu\rangle} f_a\left(\frac{\hat{r}_E}{R}\right)\approx 1 + \tau\left[2+\frac{8}{9\pi}f_a\left(\frac{\hat{r}_E}{R}\right)\right] , \label{averagemusquaredestimate}
\end{equation}
where
\begin{equation}
f_a(\rho)\equiv\rho^2\int\limits_0^\infty\frac{\mathrm{d}\zeta\, J_1^2(\zeta/\rho)}{\zeta(1+\zeta+2\zeta^2)}\approx\frac{\ln\rho}{8}; \label{fadef}
\end{equation}
the approximation is valid for large $\rho$. 

At $D=100\,\mathrm{Mpc}$, $\hat{r}_E/R\approx 2\times (10^4 - 10^{3})$ for solar-mass lenses and source star radius $R=(10-100)\,R_\odot$, 
typical for red giants which normally dominate SBF signal. Thus, the first term in~(\ref{DhatDotherrelationship}) is much greater than the second for all sensible
values of parameters. This supports the claim of the paragraph following that equation, and we will neglect the second term hereafter.

The only conceivable situation where the correlation term is non-negligible is if the microlenses are much more massive than the Sun. Since the average 
number of stars within the Einstein ring of the lens is proportional to the mass of the star, we see taking into account~(\ref{PpsiSdeltamuNbarestimate}) 
and the estimate for the luminosity ratio, that a lens mass of order $(10^3 - 10^5)\,M_\odot$ is needed for the correlation term to become of order $\tau$. 
Increasing the mass would also affect the average in~(\ref{averagemusquaredestimate}) but the dependence is much weaker -- nearly logarithmic -- in this case.

When the correlations are important the whole methodology of the SBF method breaks down as the dispersion of the pixel luminosity is no longer proportional 
to the average. Such change would manifest itself in the statistics of the surface brightness fluctuations generally enhancing the probability of large 
deviations from the mean value. Careful study of the observed fluctuation statistics therefore presents a potential means to detect a population of massive 
compact lenses. However, results of the next subsection show that the effect is unlikely to be important and we therefore do not extend our study beyond
the present observation.

\subsection{Corrections in cosmological settings}

With interstellar correlations seemingly negligible, we can focus on a single quantity 
\begin{equation}
\frac{\langle\mu^2\rangle_\phi}{\langle\mu\rangle}=1+2\tau+\int\mathrm{d}^2{\bf k}\,P_{\Delta\mu}({\bf k})|\langle\tilde\psi^2({\bf k})\rangle| +\bar{o}(\tau) \label{musquaredtomu}. 
\end{equation}
In order to evaluate this quantity three inputs are needed -- the optical depth to microlensing $\tau$ and magnification fluctuations power spectrum $P_{\Delta\mu}({\bf k})$, given by~(\ref{taudef}) and~(\ref{Pmudef}),
and the squared-luminosity-weighed average profile $\langle\tilde\psi^2({\bf k})\rangle$ defined by~(\ref{averagepsisquared}).

The former two quantities depend on the global properties of the lensing population and include underlying cosmology through the use of angular diameter
distances. In order to evaluate them, the approach of a classic paper by \citet{pressandgunn} will be used. For reasons of the economy of hypotheses we
also assume that the number density of compact lenses is constant in comoving coordinates and all lenses have the same mass $M$ while for the evaluation of distances the now-standard cosmographical model with $\Lambda=1-\Omega=0.74$ and a flat geometry will be used.

The angular diameter distance $D$ to redshift $z$ is then $D(z)=c r(z)/H_0(1+z)$, where $H_0$ is the present-day Hubble constant and
\begin{equation}
r(z)\equiv\int\limits_0^z\frac{\mathrm{d}z'}{\sqrt{\Omega(1+z')^3+1-\Omega}} . \label{rzdef}
\end{equation}
If the present-day value of the critical density in compact objects is $\Omega_d$ and it increases into the past $\propto (1+z)^3$, one has ($D_{LS}=D_S-D_L(1+z_L)/(1+z_S)$)
\begin{equation}
\tau(z_s)=\frac{3}{2}\Omega_d\int\limits_0^{z_s}\frac{\mathrm{d}z\,(1+z)r(z)}{\sqrt{\Omega(1+z)^3+1-\Omega}}\left[1-\frac{r(z)}{r(z_s)}\right] \label{taucosmological}
\end{equation}
while the magnification fluctuation power spectrum
\begin{eqnarray}
\lefteqn{P_{\Delta\mu}({\bf k})=3\Omega_d r^2_{H}\int\limits_0^{z_s}\frac{\mathrm{d}z\,(1+z)^2[r(z_s)-r(z)]^2}{(1+z_s)^2\sqrt{\Omega(1+z)^3+1-\Omega}}   }\label{PDeltamucosmological} \\
& & \hspace{16mm} \times\left\{1+k^2r^2_{H}\frac{(1+z)r(z_s)}{(1+z_s)^2 r(z)}[r(z_s)-r(z)]\right\}^{-1}\nonumber
\end{eqnarray}
where the approximation~(\ref{nuadef}) has been used and the cosmological Einstein radius $r_H$ introduced as
\begin{equation}
r_H=\sqrt{\frac{4GM}{cH_0}}.\label{rHdef}
\end{equation}
The integrals above need to be calculated numerically.

The only ingredient left is the average source star profile $\langle\tilde\psi^2({\bf k})\rangle$. As the contribution of each star is proportional
to the square of its luminosity, it is even more biased towards highest-luminosity giant stars than $\bar{l}$. However, the contribution of smaller
stars here cannot be neglected because $P_{\Delta\mu}(k)$ decreses slowly with $k$ and small stars have more extended $\tilde\psi$ profiles. We therefore
need to consider a model population of stars and find an average profile in this way.

We will assume that all stars are Lambert discs so that their spatial profiles are described by~(\ref{Lambertdisctildepsi}) and only differ in $R$, which is 
independent of the band; we do not consider any limb darkening in this simple study. Luminosity, on the contrary, depends on the wavelength, and therefore
$\langle\tilde\psi^2({\bf k})\rangle$ differs from one spectral band to another which makes $\langle\mu^2\rangle_\phi$ depend on the band, too. Since
gravitational lensing itself is achromatic, dependence via profile is the only way in which spectral bands enter $\langle\mu^2\rangle_\phi$.

\begin{figure}
\hspace{0cm} 
\includegraphics[width=80mm, angle=0]{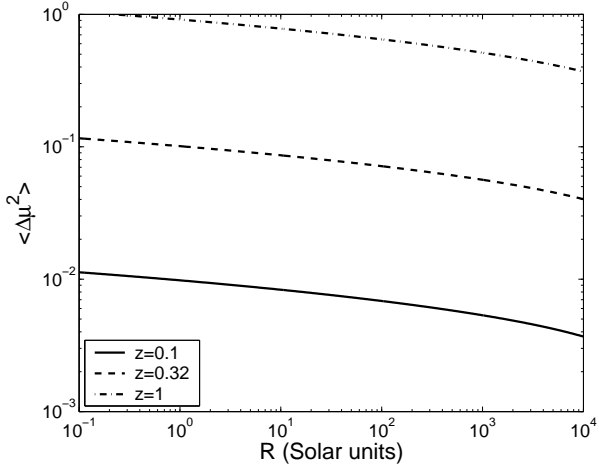}
\caption{Magnification fluctuation dispersion as a function of stellar radius (in solar units) for different redshifts. The values are computed
for a cosmological model described in the text and $r_H=8.6\times 10^{16}\,\mathrm{cm}$ corresponding to a lens mass of $1\,M_\odot$.}
\label{sigma2Rfig}
\end{figure}

To compute $\langle\tilde\psi^2({\bf k})\rangle$ we use the Bag of Stellar Tracks and Isochrones (BaSTI) population synthesis software \citep{basti, bastiweb, daniel}\footnote{The code is available at {\it http://www.te.astro.it/BASTI/}} 
In addition to stellar luminosities in nine bands -- {\it U, V, B, R, I, J, H, K} and {\it L} -- the code computes bolometric luminosities
and effective temperatures allowing us to calculate their radii using Stefan-Boltzmann law. The code has been used to compute SBF amplitudes in the infrared 
and was shown to agree with {\it 2MASS} data for Magellanic Clouds \citep{gonzalez}.

For the economy of hypotheses we have chosen to use Galactic bulge star formation history to represent that of the target galaxy stellar population. A number
of simulations with different initial seed values were performed for about a million stars each. They do not show any noticeable difference in terms of 
$\langle\mu^2\rangle_\phi/\langle\mu\rangle$ from each other. As a comparison reference, the same values for the Galactic disc-like populations were computed.

Since both $\tau$ and $\langle\sigma_\mu^2\rangle_\phi$ are proportional to the density of the microlenses $\Omega_d$, we present our results in terms of 
the correction $q\equiv(2\tau+\langle\Delta\mu^2\rangle_\phi)/\Omega_d$ defined such that the true luminosity distance to the source galaxy $d$ and its estimate $\hat{d}$ are
related by a form of~(\ref{DhatDotherrelationship}):
\begin{equation}
\hat{d}^2(z)=\frac{d^2(z)}{1+q(z)\Omega_d} +\overline{o}(\tau) \label{DhatDusingq}.
\end{equation}

Note, that when the microlenses represent a dominant fraction of the Universe matter density, the correction defined above is the one relative 
to the so-called `empty beam' distances (Dyer \& Roeder 1972; 1973). The combined action of all
the masses in a respective `filled beam' would produce an average magnification of $(1+4\tau)$ and therefore correction to the `filled beam' distance is
$q-2\tau=\langle\Delta\mu^2\rangle_\phi$. The difference between the two values is less than twenty per cent in the situations considered below.

To calculate $q$, we first compute the dispersion of the magnification as a function of stellar radius for a given source redshift $z_s$ using 
approximation~(\ref{PDeltamucosmological}) and source profile~(\ref{Lambertdisctildepsi}) -- figure~\ref{sigma2Rfig} shows examples of such functions -- 
and then average it using results of numerical simulations separately in each spectral band. Then, the $2\tau$ value is added.

\begin{figure}
\hspace{0cm} 
\includegraphics[width=80mm, angle=0]{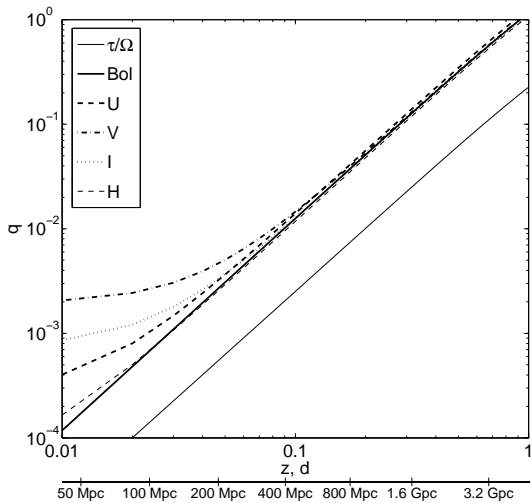}
\caption{Correction $q$, defined by~(\ref{DhatDusingq}) and the optical depth $\tau/\Omega_d$ as a function of redshift $z$
and the (true) luminosity distance $d$. Only the bolometric value and four filters -- {\it U, V, I} and {\it H} are shown to avoid crowding in the graph. A spatially flat ($\Omega+\Lambda=1$) Universe with $\Omega=0.26$ is assumed and $H_0=73\,\mathrm{km}\cdot\mathrm{s}^{-1}\cdot\mathrm{Mpc}^{-1}$ are adopted.}
\label{qfig}
\end{figure}

Figure~\ref{qfig} shows the results of our calculations. Correction $q$, defined above, and the optical depth $\tau/\Omega_d$ are shown as functions
of the redshift $z$ and true luminosity distance $d$ (upper and lower horizontal scales, respectively). We adopted the currently favoured spatially 
flat Universe for this calculation with parameters $\Omega=0.26$ and $H_0=73\,\mathrm{km}\cdot\mathrm{s}^{-1}\cdot\mathrm{Mpc}^{-1}$ \citep{cosmo06}, 
as well as a microlens mass of $1\,M_\odot$. An old, Galactic bulge-like population has been used to calculate the microlensing correction. We reiterate 
that the finite-sample effect discussed at the end of Section 2 should be considered for each galaxy individually, when the effect is computed for 
applications to real observations; it further reduces the magnitude of the lensing correction to SBF-based distances and depending on the number 
of stars in the target galaxy, this reduction can be a factor of a few for smaller galaxies.

As shown in the previous subsection, typical Einstein radii of the lenses are much larger than the stellar 
radii at most of the distances of interest and therefore for lenses of much larger mass the relation~(\ref{averagemusquaredestimate}) in the large $\rho$
limit of~(\ref{fadef}) holds; at the same time, $\hat{r}_E\propto\sqrt{M}$. Therefore, higher masses would produce corrections that are greater by 
an additive term of roughly $\tau\ln{(M/M_\odot)}/(18\pi)$, which is just 12 per cent of the $2\tau$ value for lens masses as large as $10^6\,M_\odot$. 

In order to avoid crowding in the graph, only bolometric, ultraviolet ({\it U}), visual ({\it V}) and two infrared ({\it I} and {\it H}) values of $q$ 
are depicted. In fact, {\it B} values are very close to those for {\it V}, being about five per cent lower than the latter for $z\la 0.08$ and slightly 
higher (by up to four percent) for larger redshifts. {\it K} and {\it L} graphs would be indistinguishable from those of {\it H} on the scale of this 
graph, while {\it R} and {\it J} take values roughly half-way between {\it V} and {\it I}, and {\it I} and {\it H}, respectively.

\begin{figure}
\hspace{0cm} 
\includegraphics[width=80mm, angle=0]{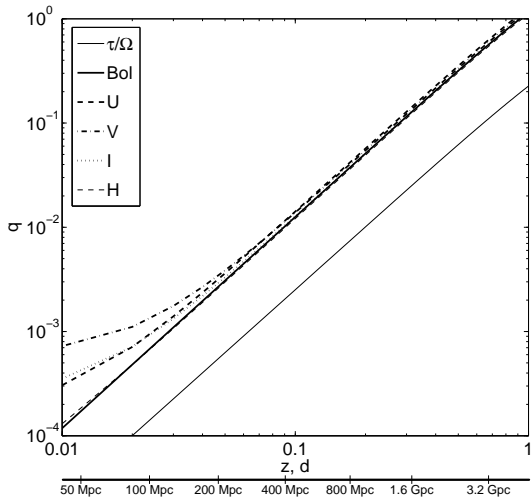}
\caption{Same as figure~\ref{qfig} computed for a Galactic disc-like stellar population. The two graphs are very similar.}
\label{qlfig}
\end{figure}

Figure~\ref{qlfig} shows the same quantities obtained for a disc-like population. One can see, that unlike the SBF fluctuation amplitude, 
the magnification correction value $q$ is very weakly dependent on the stellar population. The comment above concerning the finite-sample effect 
can be repeated in full here.

The complex behaviour of the graph in terms of the dependence on colour at low redshifts can be explained by the non-linear form of 
$\langle\Delta\mu^2\rangle(R)$ dependence.
It is proportional to the logarithm of $\hat{r}_E/R$ for small source radii and falls of as $1/R^2$ for much larger sources, 
which introduces an extra weighing factor in averaging for close galaxies, when a significant fraction of the optical depth 
is contributed by nearby lenses. This behaviour is less pronounced for the disc-like population of figure~\ref{qlfig}.

By comparing the average values of $q$ with $\langle\Delta\mu^2\rangle(R)$ in figures~\ref{qfig} and~\ref{sigma2Rfig} we see that the typical effective radius of stars that contribute
most to the correction is of order ten solar radii, except short-wavelength bands {\it U, B, V}, where one can see a significant contribution due to stars
with radii of solar order. Therefore, as is the case with the surface brightness fluctuations amplitude, the correction
is dominated by late-type giant stars. This is quite natural because the average value we are interested in is weighed with the square of stellar
luminosities, making red giants the most important contributors -- especially, for old stellar populations such as the one used in our calculations.

\section{Conclusions}

In the present paper we have considered the correction to distances inferred from the surface brightness fluctuations analysis,
caused by a potential population of compact objects along the line of sight to the source galaxy. Using a simple approximation
for microlensing magnification fluctuations we have studied how this effect can be taken into account and whether such accounting is needed in observations. Although we have considered microlensing as the main factor of individual star brightness fluctuations, the technique developed
should be applicable to a wider range of geometrical variability.

By combining the results shown in figure~\ref{qfig} with the upper limit of $\sim 0.3$ on the total fraction of matter in the energy budget of the
Universe, one can see that microlensing can for now be safely neglected when interpreting SBF-based distances. The present-day observational facilities
limit the range of the method to less than 200 Mpc while the uncertainty of this method is presently between five and ten per cent. The finite-sample 
effect discussed at the end of Section 2 further reduces the magnitude of the lensing correction.

However, for the next generations of telescopes, this situation can -- at least, in principle -- change. The main factor limiting the accuracy
of the surface brightness fluctuations method is the photon noise, which is of geometrical origin. Therefore, a naive expectation would be that
increasing the diameter of the telescope $n$-fold would lead to a directly proportional increase in the range in terms of the comoving distance $c/H_0 r(z)$. Taken with the estimates given in this study, this implies that microlensing might just start to matter when space telescope diameters get closer
to a ten-metre scale.

Alternatively, one might notice that $q$ turned to be of the same order as the total optical depth. Therefore, any large overdensity of compact microlenses -- such as those potentially associated with galaxy clusters -- would introduce noticeable change in the surface brightness fluctuations
statistics and can thus be detected in this manner. However, the properties of strong lensing mapping need to be known in full detail for such project
since local variations of the apparent luminosity distance are greatest near the critical curves of the mapping which yield galaxies most suitable
for SBF studies. Temporal variability, therefore, seems to be a better tool to study compact dark matter in galaxy clusters \citep{ourcluster}.

We conclude that microlensing by a cosmological population of compact objects is unlikely to significantly bias determination of distances using surface
brightness fluctuations technique. Similarly, present accuracy of this method does not allow to put useful constraints on the properties of such
population from the absence of systematic discrepancies between distances determined via SBF and other methods. The effect considered in this paper, 
however, should not be overlooked when the accuracy and range of the SBF method reach the level of a few per cent and a gigaparsec, respectively.

{\it Acknowledgements.} AVT is supported by IPRS and IPA from the University of Sydney. We thank Santi Cassisi, Daniel Cordier and the BaSTI team 
for performing population synthesis simulations for our study. We thank the referee for their suggestion to extend the applications of this work and
numerous useful comments.

\label{lastpage}
\end{document}